\documentclass[aps,reprint,prl,superscriptaddress]{revtex4-1}
\pdfoutput=1
  \usepackage{ulem}
  \usepackage{graphicx}
\usepackage{xcolor}
\usepackage{hyperref}

\newcommand{\etal}{{\it et al.}}

\newcommand{\basOrig}{LQC}
\newcommand{\basCut}{LQC-c}
\newcommand{\basOrigEmb}{LQC-0}
\newcommand{\basCutEmb}{LQC-0c}
\newcommand{\basOcrys}{TZVP-a}
\newcommand{\basOskr}{TZVP-b}

    \def\CSs{chemshifts}
  \def\CTEP{compound-tunable embedding potential}
  \def\abinitio{{\it ab~initio}}
  
  \def\dfel{$d/f-$}

\begin{document}
\title{Compound-tunable embedding potential method and its application to fersmite crystal}
\author{D.A.\ Maltsev}\email[]{malcev\_da@pnpi.nrcki.ru}
\affiliation{Petersburg Nuclear Physics Institute named by B.P.\ Konstantinov of National Research Center ``Kurchatov Institute'' (NRC ``Kurchatov Institute'' - PNPI), 188300, Russian Federation, Leningradskaya oblast, Gatchina, mkr.\ Orlova roscha, 1.}
\author{Yu.V.\ Lomachuk}
\affiliation{Petersburg Nuclear Physics Institute named by B.P.\ Konstantinov of National Research Center ``Kurchatov Institute'' (NRC ``Kurchatov Institute'' - PNPI), 188300, Russian Federation, Leningradskaya oblast, Gatchina, mkr.\ Orlova roscha, 1.}
\author{V.M.\ Shakhova}
\affiliation{Petersburg Nuclear Physics Institute named by B.P.\ Konstantinov of National Research Center ``Kurchatov Institute'' (NRC ``Kurchatov Institute'' - PNPI), 188300, Russian Federation, Leningradskaya oblast, Gatchina, mkr.\ Orlova roscha, 1.}
\affiliation{Saint Petersburg State University, 7/9 Universitetskaya nab., 199034 St. Petersburg,  Russia}
\author{N.S.\ Mosyagin}
\affiliation{Petersburg Nuclear Physics Institute named by B.P.\ Konstantinov of National Research Center ``Kurchatov Institute'' (NRC ``Kurchatov Institute'' - PNPI), 188300, Russian Federation, Leningradskaya oblast, Gatchina, mkr.\ Orlova roscha, 1.}
\author{L.V.\ Skripnikov}
\affiliation{Petersburg Nuclear Physics Institute named by B.P.\ Konstantinov of National Research Center ``Kurchatov Institute'' (NRC ``Kurchatov Institute'' - PNPI), 188300, Russian Federation, Leningradskaya oblast, Gatchina, mkr.\ Orlova roscha, 1.}
\affiliation{Saint Petersburg State University, 7/9 Universitetskaya nab., 199034 St. Petersburg,  Russia}
\author{A.V.\ Titov}\email[]{titov\_av@pnpi.nrcki.ru}
\affiliation{Petersburg Nuclear Physics Institute named by B.P.\ Konstantinov of National Research Center ``Kurchatov Institute'' (NRC ``Kurchatov Institute'' - PNPI), 188300, Russian Federation, Leningradskaya oblast, Gatchina, mkr.\ Orlova roscha, 1.}
  \date{\today, version: `EP-fersmite-42dm'}
\begin{abstract}
 Compound-tunable embedding potential (CTEP) method is proposed. 
A fragment of some chemical compound, ``main cluster'' in the present paper, is limited by boundary anions such that the nearest environmental atoms are cations. The CTEP method is based on constructing the embedding potential as linear combination of short-range ``electron-free'' spherical  ``tunable'' pseudopotentials
for cations from nearest environment,
 whereas the long-range CTEP part consists of Coulomb potentials from optimized fractional point charges centered on both environmental cations and anions.

A pilot application of the CTEP method to the fersmite crystal, CaNb$_2$O$_6$,
is performed and a remarkable agreement of the electronic density and interatomic distances within the fragment with those of the original periodic crystal calculation is attained. 
Characteristics of ``atoms-in-compounds'' 
 [A.V.\ Titov \etal, Phys.Rev.A {\bf 90}, 052522 (2014)]
 which are of great importance for materials containing $f$- and $d$-elements (Nb in fersmite) are considered on examples of 
chemical shifts of $K_{{\alpha}_{1,2}}$ and $K_{{\beta}_{1,2}}$ lines of X-ray emission spectra in niobium.
A very promising potential of this approach in studying variety of properties of point defects containing $f$- and heavy $d$-elements with relativistic effects, extended basis set and broken crystal symmetry considered is discussed.
\begin{figure}[h]
	\centering
	\includegraphics[width=0.8\linewidth]{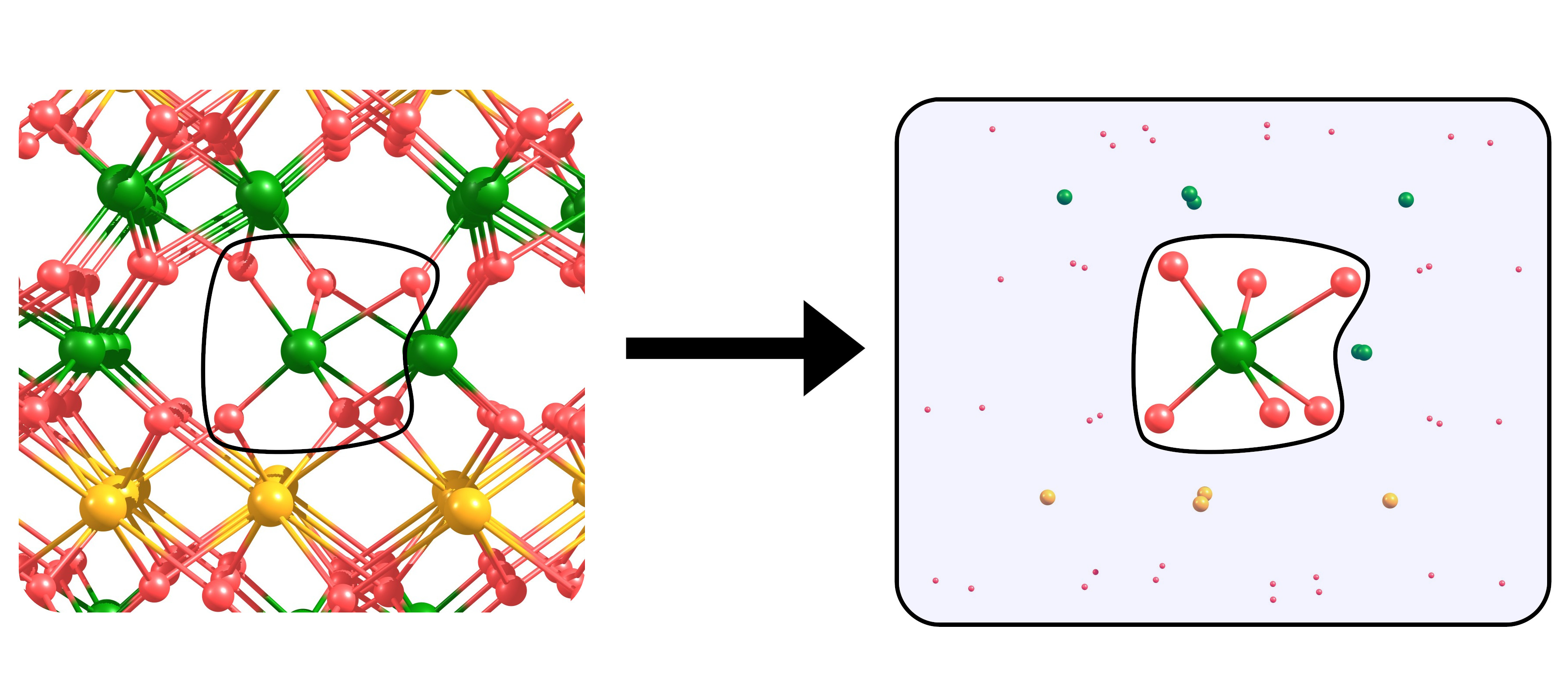}
\end{figure}
 
   \end{abstract}
 
\maketitle

\section{Introduction}
\label{Intro}

Impressive recent achievements in creating experimental facilities to study local atomic-scale electronic structures in material science like X-ray free-electron lasers, synchrotrons \cite{Chergui:16_StrDin}, high harmonic generation sources \cite{Ghimire:19NatPhys}, etc.\ open new era in investigating  materials and defects containing heavy transition metals ($d$-elements), lanthanides and actinides ($f$-elements);
we will 
  designate all these atoms as \dfel elements below.
However, theoretical possibilities in direct studying electronic structures on atomic-scale or, by other words, properties of 
atoms-in-compounds (AiC) 
\cite{Titov:14a, Skripnikov:15b, Lomachuk:13} are yet hampered by several challenges in quantum chemical description of such systems. They first include necessity of highest-level treatment of relativistic and correlation effects simultaneously. Besides, polyvalent \dfel elements often have pronounced multireference character and high density of low-lying electronic states. As a result, opportunities for direct \abinitio\ study of materials containing \dfel elements with required accuracy can be blocked by unacceptable computational costs (see analysis for ThO in \cite{Skripnikov:15a}).
An alternative way to explore such a material is to reduce its studying to a molecular-type investigation of some its fragment, assuming that relaxation of the rest of the crystal (environment of the fragment) in processes under consideration is negligible. In this case one can consider influence of the environment on the fragment by some approximate {\it embedding potential} 
  to improve the quality
of description of phenomena localized on the fragment
  using extended possibilities of its studying by molecular methods. 
Such fragment with embedding potential
is usually called the ``embedded cluster'' or ``cluster, embedded in a crystal''.

The embedding potential theories are based on the idea of freezing the external environment of the embedded cluster; they are conceptually similar to the effective core potential (pseudopotential or PP)  theories which are originated by the frozen core approximation in atoms. Such theories are highly demanded for studying point defects, localized properties and processes in solids and other many-atomic systems, particularly, if they contain 
\dfel elements.
Among the phenomena of increasing interest, one could highlight opening new possibilities to study
magnetic structure and valence state of \dfel elements in materials \cite{Lomachuk:19a},
localized excitations in crystals and matrices (molecular rotors \cite{Comotti:16_ACR}, chimeras and intrinsic localized modes (discrete breathers) \cite{Dmitriev:16}, electronic transitions and magneto-optical effects in point defects \cite{Wickenbrock:16_NVc}), new physics in ferroelectrics (see \cite{Skripnikov:16a} and refs.), etc.
Description of electronic structure in many-atomic systems containing \dfel elements, particularly, within periodic models, is problematic to-date. Excitation energies for valence electrons in \dfel elements can be very small, within errors of density functional theory (DFT) approximation. The situation is most difficult for light actinides, which show both lanthanide-like and transition-metal-like behavior. Therefore, calibration of the exchange-correlation DFT functionals should be done to choose its appropriate version, which provide correct valence state of \dfel elements in a compound under consideration. 
Note that calculation of a crystal fragment of small size (including up to $\sim$10 atoms, with a central \dfel element and its first anionic coordination sphere) using the embedding potentials and combining \cite{Zaitsevskii:13b, Skripnikov:16a} advanced two-component (relativistic) versions of density functional  \cite{vanWullen:02} and coupled-cluster (see \cite{Skripnikov:15a, Pasteka:17_RCC-Au} and refs.) 
theories can be done in practice. 
Furthermore, one can perform relativistic calculation 
of a larger cluster with the chosen DFT functional 
with impurity actinides and vacancies to take relaxation of its neighbors into account. The relativistic coupled-cluster (RCC) corrections to 
the DFT calculation of the cluster can also be done (see pilot combined study in \cite{Skripnikov:16a}).

The embedding potentials are actively used in solid-state studies (see review  \cite{Abarenkov:16} and refs). We mention here two of recent developments of embedding theories. In series of papers (see refs.\ in \cite{Wouters:17_dmet}), density matrix embedding theory (DMET) is developed. It describes a finite fragment within the surrounding environment
such that the local density of states can be obtained when working with a Fock space of ``bath'' (environmental) orbitals.

Going beyond the self-consistent field (SCF) approximation for environment, H\"ofeneer with colleagues considered DFT-based
methods which encompasses wave-function theory-in-DFT (WFT-in-DFT) and the DFT-based subsystem formulation of response theory (DFT-in-DFT) \cite{Hoefener:12_emb}.
These approaches allow one to take account of the environmental relaxation due to perturbation by a point defect in a small fragment.

Though such combining procedures are quite natural from theoretical  point of view, in practice they are not trivial and were applied to relatively simple compounds containing only light atoms. 
They are rather expensive computationally for the systems containing \dfel elements which are of interest in practice.

In turn, the pseudopotential technique has proved to be quite successful and universal toolkit to combine WF-based frozen core approximation with DFT treatment of electronic structure in the valence region of a compound. Besides, relativistic spin-dependent, quantum electrodynamics (QED) and correlation effects can be taken into account with high level of accuracy within the PP approximation and only in those regions (core or valence) in which they are important. In particular, core electronic states can be frozen as atomic spinors, whereas the valence ones may be treated as spin-orbitals (with the spin-orbit corrections often taken on the last computational stage only) \cite{Titov:99}. Even large-core PPs can handle relaxation (``response'') of atomic electronic structures caused by small external perturbations appropriately if they are ``transferable''
  \cite{Goedecker:92a, Fromager:04} and
generated for appropriate {\it effective states of the atoms in a compound} \cite{Titov:14a}. Therefore, such large-core PPs can be used to describe environmental atoms.

In the given paper, a new method, {\it \CTEP} (CTEP), is proposed to describe local properties and 
processes in minerals, particularly, if the mineral contains point defects with \dfel elements. The capability of the method is demonstrated here on example of the fersmite crystal, CaNb$_2$O$_6$, as a representative of tantalum-niobate minerals' group.
The corresponding CTEP for xenotime (yttrium orthophosphate mineral, YPO$_4$, having both ionic and covalent bonds)  is discussed in \cite{Lomachuk:19a} and applied there to study thorium and uranium containing point defects. The other application of CTEP,
to the structures containing periodically arranged lanthanide atoms (which can have open 4f shell), is considered in \cite{Shakhova:19a} on examples of YbF$_2$ and YbF$_3$ crystals.

\section{Method}

In the framework of CTEP version of the embedding potential theory, the following procedure for calculating the electronic structure of a crystal fragment 
which can include a point defect is implemented:

 (1) High-level periodic DFT calculation of a crystal without point defects.
 
 (2) Cutting a fragment out of a crystal with a central metal atom (which can be further replaced by a vacancy, impurity atom, etc.) and its nearest anionic environment. Thus, the first coordination sphere consists of a small number of atoms that is usually not more than 12 (as in the case of dense packing). This structure will be referred further as the ``main cluster''.
  
(3) For the main cluster one can choose a ``nearest cationic environment'', NCE, from the lattice atoms (second coordination sphere, etc.) and a ``nearest anionic environment'', NAE, including a set of all anions 
which are nearest to the NCE atoms except those of the main cluster. The main cluster together with the nearest environment, i.e.\ main cluster + NCE + NAE (or main cluster + CTEP), will be referred as the {\it extended cluster}, or just a cluster.

The NCE is described by means of nonlocal (semi-local) ``electron-free'' (or ``largest-core'') PPs for cations $Cat_i$, 0ve-PP/$Cat_i^{+n_i}$, which are 
generated for 
the effective states of the cations in the given crystal such that:  
(i) 
all the one-electron states (orbitals or spinors) occupied for the oxidation state $+n_i$ of the $Cat_i$ atom in the crystal are treated as core ones within the 0ve-PPs;  
(ii) 
the addition of such 0ve-PPs 
                     does not change the number of electrons in calculation of the extended cluster compared to the main cluster.

Each NCE atom is described by the PP and fractional point effective charge. ``Tuning'' the 0ve-PPs 
for environmental cations is carried out self-consistently in 
the periodic DFT calculations of pure crystal 
to minimize displacement forces on $Cat$'s and other atoms in the unit cell
at the optimized DFT geometry of the crystal obtained in step (1).

The nearest anionic environment is described by only the Coulomb potentials of the effective charges. 
Both the anionic (NAE) and cationic (NCE) charges located at the lattice sites are optimized when constructing the CTEP to reproduce the spatial structure
of the main cluster as a fragment from the periodic study (at step 1) in the molecular-type calculation of the extended cluster with only point symmetry of the crystal fragment taken into account.
Our charge optimization criterion is minimization of sum of squares of forces on atoms from the main cluster, with the constraint that the total charge of the extended cluster is fixed to zero.
Note, that relaxation of electronic structure in both main cluster and nearest environment regions is taken into account within CTEP when one
considers
processes and point defects localized on the main cluster despite the coordinates of the environmental atoms are not changed.

\section{Computational details}

All calculations were performed using DFT method with PBE0 functional. The solid-state and cluster calculations were performed with {\sc crystal-17} \cite{crystal:17} and relativistic molecular DFT \cite{Wullen:10} packages, respectively.

The core pseudopotentials generated by our group \cite{Titov:99} for the Ca and Nb atoms were used. 
The original basis sets, corresponding to these PPs (mentioned below as \basOrig{} --
 (8,8,7,2)/[6,6,4,2]
 for Nb and
(5,5,4,1)/[5,5,4,1]
  for Ca) were cut and contracted for use in solid-state calculations (mentioned below as \basCut{} --
 (5,5,5,2)/[3,3,2,2]
  for Nb and 
 (4,3,3,1)/[3,3,3,1]
 for Ca).

The tuned 0ve-PPs with the basis sets combined from the valence orbitals of the original basis sets (\basOrig{} and \basCut{})
and core exponents of the pseudo-orbital expansion generated in the present work
were used for the Ca and Nb ``pseudoatoms'' treated as NCE.
Further, we will refer to 
these
basis sets as \basOrigEmb{}, and \basCutEmb{}, respectively.

The O\_pob\_TZVP\_2012 basis set designated below as \basOcrys{} was used for oxygen atoms.
For a comparison, calculation of the Nb-centered cluster, along with untruncated (\basOrig{} and \basOrigEmb{}) versions of Ca and Nb basis sets, was carried out with an augmented version of TZVP basis for oxygen (introduced in \cite{Lomachuk:18en}) and mentioned below as \basOskr{}
(12,7,2)/[6,4,2].

  Since the main goal of the paper is to examine the quality of the cluster simulation of periodic structures within the CTEP approximation, all our cluster calculations (except evaluation of XES \CSs{}) were performed without spin-orbit interaction for consistency with periodic calculations using {\sc crystal} package, in which only scalar-relativistic effects can be taken into account due to the software limitations. Influence of spin-orbit effects is discussed in more details in papers \cite{Lomachuk:19a, Shakhova:19a}.

\section{Results and discussions}

\textbf{1. Periodic calculations}

The fersmite crystal belongs to {\it Pcan} space group and consists of five non-equivalent atomic types: Ca, Nb and three different O types.  Both atomic positions and cell parameters were optimized with only crystal symmetry group fixed. The resulting structure was close to the experimental one with the average bond length error about 0.8\%.

The most significant difference was for the specific Ca-O bond with interatomic distance of 2.80{{\AA}} in the experimental structure {\it vs.}\ 2.71{{\AA}} in the DFT optimized one.
In both cases this interatomic distance is considerably larger than the sum of crystal radii (1.26 {{\AA}} for Ca(VIII) and 1.24 {{\AA}} for O(IV)), and
it is {\it a~priori} unclear whether the corresponding atoms should be considered as the neighbor ones or not, so that for building the 
Calcium-centered
clusters we have considered both cases of the first coordination sphere: CaO$_{6}$ and CaO$_{8}$.

\textbf{2. Cluster calculations}

\begin{figure}[h]
	\centering
	\includegraphics[width=0.99\linewidth]{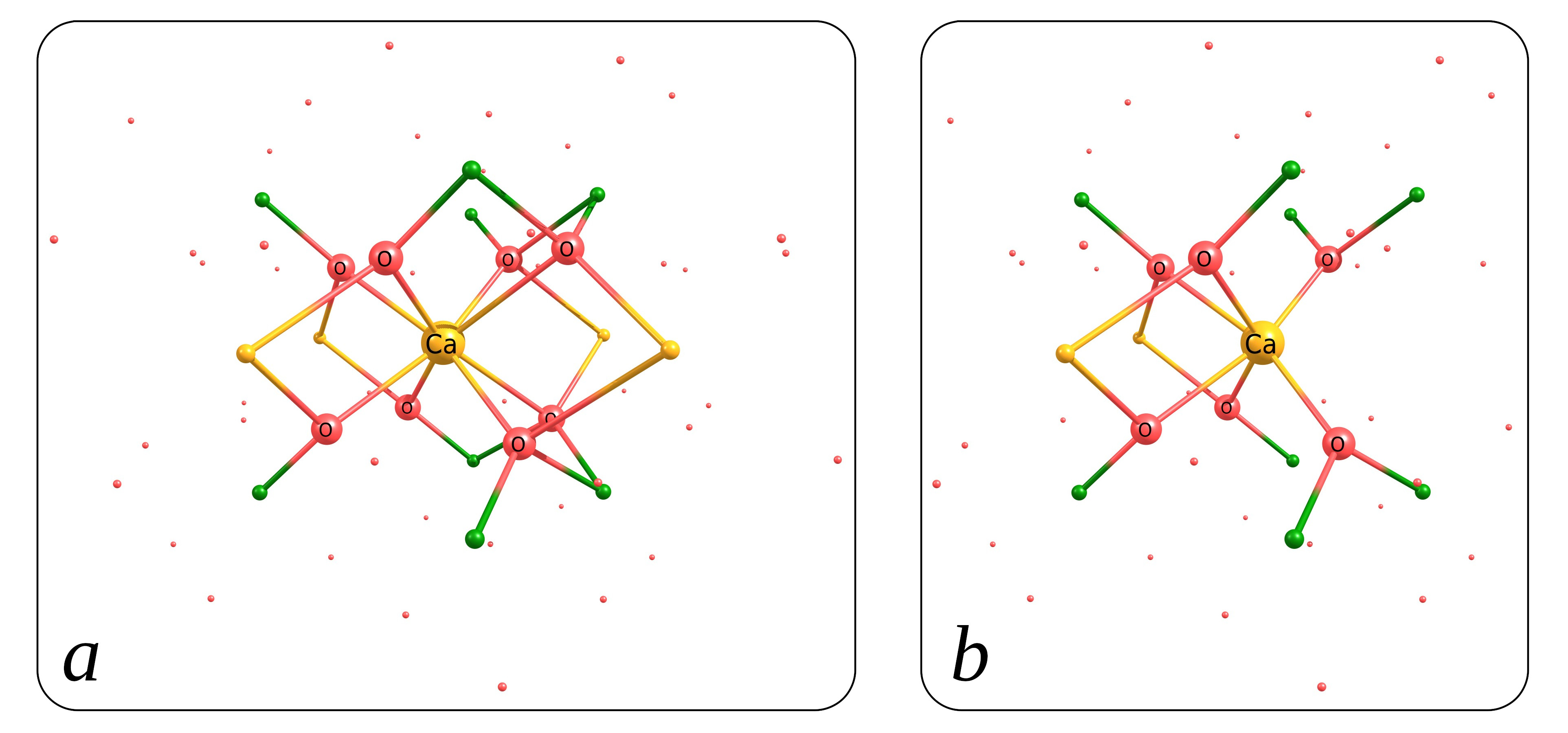}
	\caption{Ca-centered clusters:  $($a$)$ large [CaO$_{8}$][Ca$_{4}$Nb$_{8}$][O$_{44}$], and $($b$)$ small [CaO$_{6}$][Ca$_{2}$Nb$_{8}$][O$_{38}$]. 
NCE atoms are shown as spheres of half-radius without caption (of the same colour as for corresponding atoms of the main 
cluster and with the drawn bonds to the oxygen atoms) 
and NAE charges are shown as the dot-like semi-transparent spheres.
}
	\label{fig:cluster_ca}
\end{figure}

Clusters centered on both Ca and Nb cations were built together with CTEPs.
As mentioned above, the number of the nearest neighbors of Ca atom in fersmite which should be included to the main cluster is ambiguous, so, two main clusters with different CTEPs were built for the central calcium case: the small one, [CaO$_{6}$][Ca$_{2}$Nb$_{8}$][O$_{38}$] and the large one, [CaO$_{8}$][Ca$_{4}$Nb$_{8}$][O$_{44}$] (Figure \ref{fig:cluster_ca}).

\begin{figure}[h]
	\centering
	\includegraphics[width=0.99\linewidth]{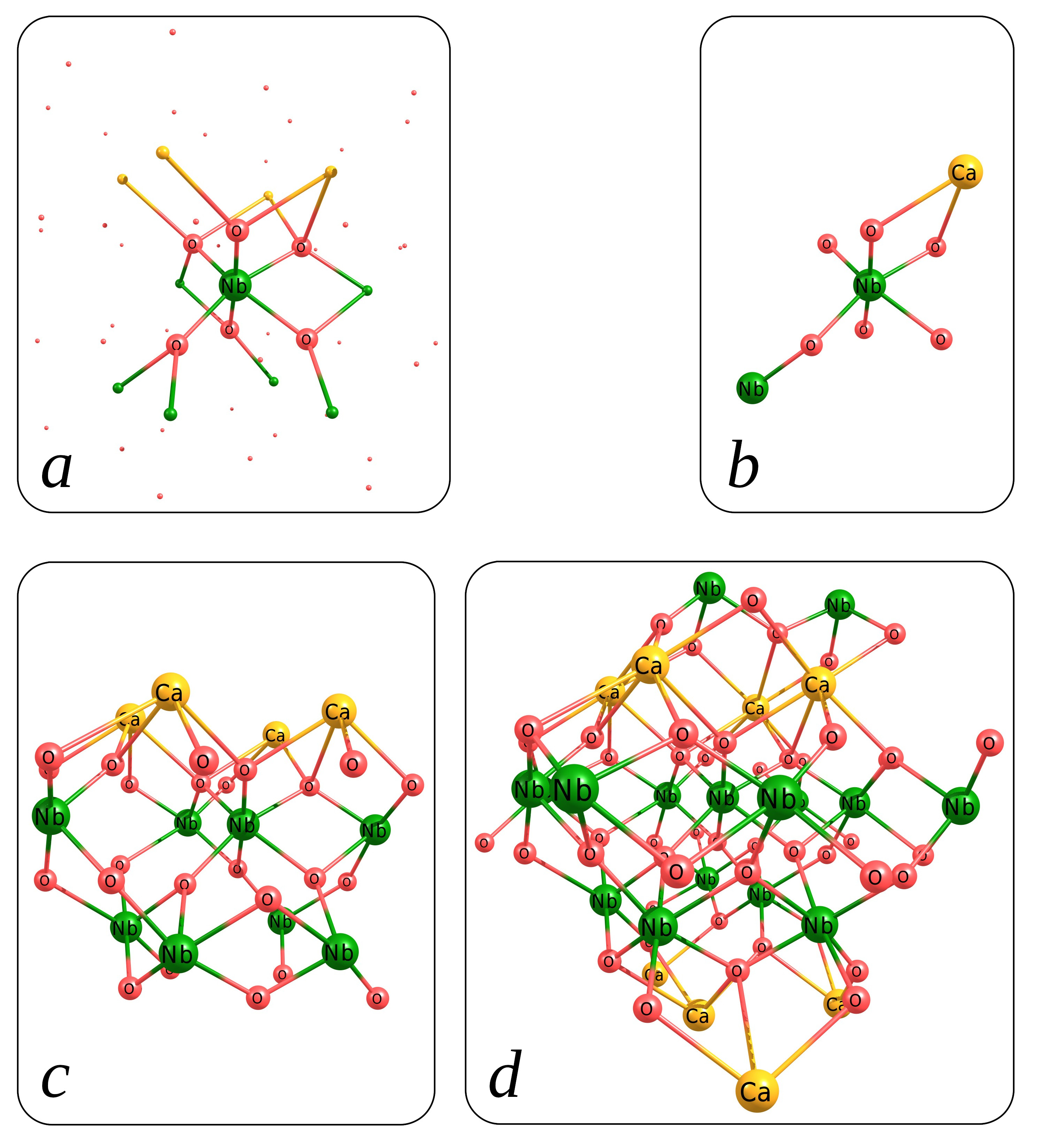}
	\caption{Nb-centered clusters: $($a$)$ Cluster with CTEP, [NbO$_{6}$][Ca$_{4}$Nb$_{6}$][O$_{41}$]; $($b-d$)$ three stoichiometric clusters with 1, 4 and 8 minimal formulas [CaNb${_2}$O${_6}$], respectively}
	\label{fig:cluster_nb}
\end{figure}

For the central Nb atom, only one cluster with CTEP, [NbO$_{6}$][Ca$_{4}$Nb$_{6}$][O$_{41}$], was built. However, for a comparison, three stoichiometric clusters for the niobium surroundings (without CTEP) were built and considered. The first one is for a single minimal formula [CaNb${_2}$O${_6}$], whereas the second and third ones are for 4 and 8 minimal formulas, respectively (Figure \ref{fig:cluster_nb}).

Despite all the atoms for the stoichiometric clusters are treated on equal footing, for convenience of comparison we will refer to the central area (NbO$_{6}$ group) as the ``main cluster''.

For each of the cluster with CTEP, the fractional charges on the cationic and anionic shells were optimized, and the resulting root mean square (RMS) forces on the main cluster were obtained. For a verification, the main cluster was re-optimized with fixed CTEP parameters and the resulting geometry was compared to the original one.

For the niobium-centered cluster with CTEP, optimization of geometry was additionally performed with three extended basis sets:
\textit{(1)} the same as for the periodic calculations except for the original uncut \basOrig{} basis on the niobium atom,
\textit{(2)} the \basOrig{} basis on the niobium atom with \basOskr{} basis on oxygen atoms, and
\textit{(3)} the \basOrig{} basis on the niobium atom with \basOskr{} basis on oxygen atoms and \basOrigEmb{} on the NCE atoms.

Additionally, optimization of geometry was also performed for the stoichiometric clusters, with the central NbO$_{6}$ being treated as the main cluster, and the remaining atoms as fixed ``embedding''.

For all the clusters the remaining forces after optimization of geometry were negligible. All the forces and displacements are listed in the Table \ref{table:clusteropt}.

\begin{table}[h!]
	\caption{Forces on the atoms of the main 
		cluster and atomic displacements within the main cluster after its optimizations}
	\label{table:clusteropt} 
	\begin{tabular}{lccc}
		\hline
		Structure & RMS force (a.u.) & RMS dispacement({{\AA}}) \\  
		\hline
		Ca-CTEP-small & 1.1$\cdot$10$^{-5}$ & 1.5$\cdot$10$^{-4}$ \\ 
		\hline 
		Ca-CTEP-large & 2.6$\cdot$10$^{-5}$ & 3.4$\cdot$10$^{-4}$ \\ 
		\hline 
		Nb-CTEP & 3.3$\cdot$10$^{-5}$ & 2.5$\cdot$10$^{-4}$ \\ 
		\hline 
		Nb-CTEP \textit{(basis set 1)}\footnotemark[1] & 3.2$\cdot$10$^{-3}$ & 1.3$\cdot$10$^{-2}$ \\ 
		\hline 
		Nb-CTEP \textit{(basis set 2)}\footnotemark[1] & 4.8$\cdot$10$^{-3}$ & 1.5$\cdot$10$^{-2}$ \\ 
		\hline 
		Nb-CTEP \textit{(basis set 3)}\footnotemark[1] & 4.9$\cdot$10$^{-3}$ & 1.6$\cdot$10$^{-2}$ \\ 
		\hline 
		Nb-stoichiometric (1x) & 8.8$\cdot$10$^{-2}$ & 2.0\footnotemark[2] \\ 
		\hline 
		Nb-stoichiometric (4x) & 5.8$\cdot$10$^{-2}$ & 8.8$\cdot$10$^{-1}$\footnotemark[2] \\ 
		\hline 
		Nb-stoichiometric (8x) & 5.0$\cdot$10$^{-2}$ & 1.9$\cdot$10$^{-1}$ \\ 
		\hline 
	\end{tabular}
	\footnotetext[1]{Extended basis sets, described
                 in subsection ``Cluster calculations''.}
	\footnotetext[2]{Structure breaks after cluster optimization resulting in decrease of niobium coordination number.}
\end{table}

For all clusters with CTEP and original basis set, the forces and displacements are small enough to allow us to assume precise reproducibility of the geometry with CTEP method. In the case of Nb cluster, expansion of basis set leads to increase of both RMS forces and displacements by 2 orders of magnitude, however, the absolute numbers are still comparable to general errors of the DFT method.
The agreement of cluster-optimized geometry with the experimental one stays on the same level as that for the crystal-optimized geometry. 

For the stoichiometric clusters RMS forces are about 3 orders larger, and only slowly decrease with the increase of a cluster size, so one can expect that it requires a much larger stoichiometric cluster to reproduce crystal structure at the same level as the CTEP method. Optimization of the stoichiometric clusters consisting of 1 and 4 formulas breaks the correct coordination number of central Nb, while the largest one (8 formulas) preserves the correct coordination
number, but the displacements of corresponding atoms are significantly larger than those for the CTEP case. 

Overall, we can state that the crystal fragment structure obtained in the periodic calculation within the chosen DFT approximation can be reproduced in the cluster calculation with CTEP without notable decrease of the accuracy compared to the periodic DFT case, while the stoichiometric cluster approach yields much larger errors.

\textbf{3. Electronic density comparison}

To estimate the reproducibility of properties in the cluster model with CTEP, electronic density cube files were obtained for the periodic crystal study and for each cluster. The cube grid was chosen to be the same in all cases with the orthogonal unit vectors of about 0.056 a.u.

As a quantitative criterion we provide the difference between 
the cluster and crystal electronic densities, 
calculated by the following formula:
\[ d(r) = \frac{1}{4\pi}\oint d \Omega \left | \rho_{cluster}(\vec{r})-\rho_{crystal}(\vec{r}) \right | \]

\begin{figure}[h]
	\centering
	\includegraphics[width=0.99\linewidth]{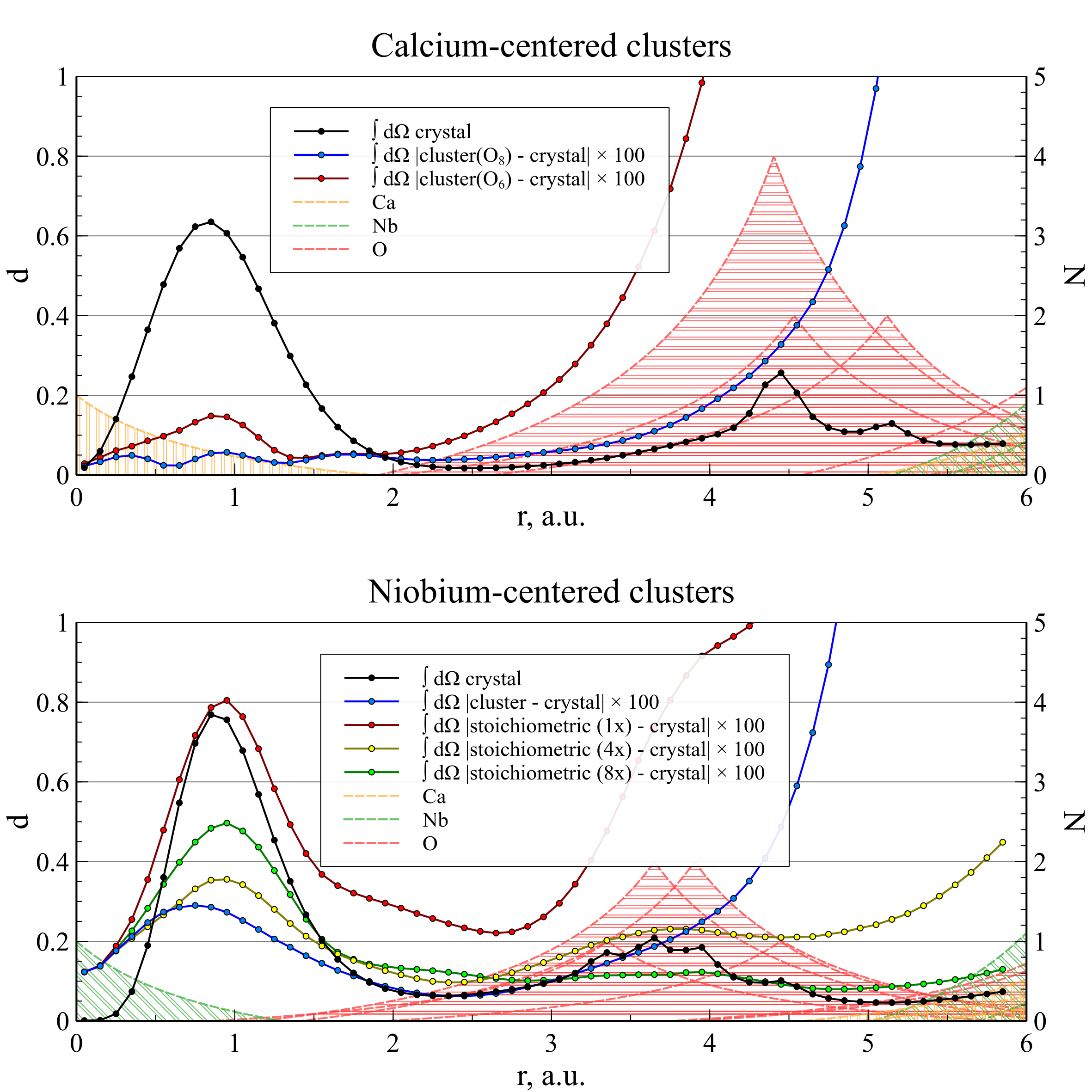}
	\caption{The radial dependence of electronic density differences for the clusters under study. Black lines represent total density. Colored solid lines correspond to the integral of absolute difference ($d(r)$), multiplied by factor of 100. Filled peaks at the bottom qualitatively represent the position of the neighbour atoms (color denotes atom type, width at the bottom is equal to crystal radius, and the peak height is proportional to the number of atoms (right vertical axis) at the same distance from the center).}
	\label{fig:graph}
\end{figure}

On Figure \ref{fig:graph} this value is plotted for all clusters under study, except those with extended basis set. The bottom dashed peaks qualitatively represent the electronic density of atoms from the main cluster. The black curve is the total density, and all the difference curves are multiplied by factor 100, so that intersections between total and difference curves correspond to the 1\% deviation.

For Ca-centered clusters the density in near surroundings of 
the 
central atom is reproduced with a good accuracy. The larger cluster (with 8 neighbors) both yields better agreement in the central area and preserves the tolerable error value in a larger radial range, which can be explained by the influence of the two oxygen atoms (at 5.1 a.u.), which are excluded from the smaller cluster.

The Nb-centered CTEP cluster yields larger error in the vicinity of the central atom, while at larger R the difference is comparable to that of the 8-neighbour Ca-centered cluster. From comparison of the embedded Nb-centered cluster with the stoichiometric ones, it follows that the embedding model reproduces electronic density within $R{<}3$~a.u.\ with notably better accuracy than any of stoichiometric clusters, while being comparable by means of computational expenses to the smallest stoichiometric one. The considerable difference in the central area is almost the same for all Nb clusters, so it is likely to result not from the CTEP inaccuracy, but from
computational features of the solid-state and cluster software (clarification of which in detail will be the subject of further research).

    At $R\,{\sim}\,4\,{\div}\,5$~a.u.\
 the difference increases greatly for all the clusters with CTEPs, which reflects 
 the matter of fact that outer core electrons of the cationic-layer atoms are included into the periodic calculations while being excluded from the cluster study using the ``electron-free'' PPs, since the CTEP model is intended to describe only localized fragments of crystals.
For both Ca-- and Nb--centered clusters with CTEP the electronic density difference lies within 2$\%$ at the effective crystal radius (2.38 a.u.\ for Ca and 1.47 a.u.\ for Nb).

\textbf{4. Chemical shifts of X-ray emission 
lines}

  Stabilization
 of computed internuclear distances with respect to a basis set enlargement 
is a good probe for the basis set saturation in the valence region of a compound when effective Hamiltonian and exchange-correlation functional are fixed. In turn, chemical shifts (\CSs{}) of lines of X-ray emission (fluorescence) spectra, XES (see \cite{Siegbahn:08, Sumbaev:78, Bokarev:18e} and references therein), are sensitive to local variation of electronic densities in the atomic core regions \cite{Lomachuk:13} that cover probing the basis set completeness in theoretical study. Moreover, the XES \CSs{} together with other AiC properties can provide a pretty informative array of data about the electronic structure near heavy \dfel atoms in a solid.
The energetic shifts of a characteristic transition between different core shells of an atom in variety of compounds allow one to explore corresponding core regions \cite{Lomachuk:18en} and study various AiC characteristics. In particular, the \CSs{} of K$_{\alpha}$-lines of \dfel elements are mainly sensitive to occupation numbers of appropriate $d$ and $f$-shells, whereas the \CSs{} of K$_{\beta}$-lines are already sensitive to distances to the ligands and their types.
It is not less important that characteristic XES lines can be easily 
identified for any atom of interest and the XES \CSs{} can be 
measured on the atoms having sufficient fraction
 in a material, 
thus providing data to characterize an effective state of {\it any}
atom for all these compounds \cite{Lomachuk:13, Titov:14a, Skripnikov:15b, Shakhova:17rad, Lomachuk:18en}.

 In Table \ref{table:chemshifts-embedding} the \CSs{} are presented for the Nb atom in the embedded cluster for four mentioned above basis sets and for the original structure ``crys'' (that is taken from periodic calculation) and re-optimized cluster structure after the CTEP construction (case ``opt'').

\begin{table}[h!]
	\caption{X-ray chemical shifts on the Nb atom in clusters with CTEP (in meV)}
	\label{table:chemshifts-embedding} 
	\begin{tabular}{llllcccc}
		\hline
                     \multicolumn{3}{c}{basis} & structure & K$_{\alpha2}$ & K$_{\alpha1}$ & K$_{\beta2}$ & K$_{\beta1}$  \\  
\cline{1-3}
		Nb       & O       & NCE &           &               &               &              &               \\  
		\hline 
	\basCut{} & \basOcrys{} & \basCutEmb{} & crys\footnotemark[1] & 345 & 364 & 227 & 246 \\
	\hline
	\basCut{} & \basOcrys{} & \basCutEmb{} & opt\footnotemark[2] & 345 & 364 & 226 & 246 \\
	\hline
	\basOrig{} & \basOcrys{} & \basCutEmb{} & crys\footnotemark[1] & 344 & 362 & 221 & 235 \\
	\hline
	\basOrig{} & \basOcrys{} & \basCutEmb{} & opt\footnotemark[2] & 341 & 359 & 213 & 227 \\
	\hline
	\basOrig{} & \basOskr{} & \basCutEmb{} & crys\footnotemark[1] & 348 & 366 & 235 & 247 \\
	\hline
	\basOrig{} & \basOskr{} & \basCutEmb{} & opt\footnotemark[2] & 346 & 364 & 236 & 248 \\
	\hline
	\basOrig{} & \basOskr{} & \basOrigEmb{} & crys\footnotemark[1] & 345 & 363 & 227 & 236 \\
	\hline
	\basOrig{} & \basOskr{} & \basOrigEmb{} & opt\footnotemark[2] & 343 & 361 & 225 & 235 \\
	\hline
	\end{tabular} 
\footnotetext[1]{Original structure (optimized in periodic calculation)}
\footnotetext[2]{The fragment structure, optimized in cluster calculation
with the corresponding basis set}
\end{table}

The dispersion of the K$_{\beta1,2}$ data is up to  ~10\,\% with increasing the basis set size that is not negligible. Thus, the corrections on incompleteness of the
basis set in the crystal calculations are highly desirable. Such corrections can be rather easily evaluated for the main cluster of minimal size (with a central atom and first coordination sphere only) in contrast to the cases of large cluster or periodic structure studies.

In Table \ref{table:chemshifts-stoichiometric} the chemical shifts are presented for the Nb atom in the stoichiometric cluster with the periodic-optimized (``original'') and cluster-optimized structures using DFT.

\begin{table}[h!]
	\caption{X-ray chemical shifts on the Nb atom in stoichiometric clusters (in meV)}
	\label{table:chemshifts-stoichiometric}
	\begin{tabular}{llcccc}
		\hline
		Cluster type\footnotemark[1] & structure & K$_{\alpha2}$ & K$_{\alpha1}$ & K$_{\beta2}$ & K$_{\beta1}$  \\  
		\hline 
		stoichiometric (1x) & crys & 341 & 360 & 218 & 238 \\ 
		\hline 
		stoichiometric (1x) & opt & 234 & 249 & 89 & 102 \\ 
		\hline
		stoichiometric (4x) & crys & 336 & 354 & 217 & 234 \\ 
		\hline 
		stoichiometric (4x) & opt & 171 & 171 & -33 & -29 \\ 
		\hline
		stoichiometric (8x) & crys & 343 & 362 & 229 & 247 \\ 
		\hline 
		stoichiometric (8x) & opt & 353 & 373 & 267 & 285 \\
		\hline
	\end{tabular}
\footnotetext[1]{The basis set for all structures corresponds to the two upper rows in Table~\ref{table:chemshifts-embedding} --- \basCut{} for all Nb and Ca atoms and \basOcrys{} for O atoms}
\end{table}

The most important result of calculations of XES \CSs{} with stoichiometric clusters is that the \CSs{} for the cluster with even 8 formulas cannot be considered as converged ones for K$_{\beta}$ \CSs{} to those of periodic structure 
(Table~\ref{table:chemshifts-embedding}) 
in contrast to those for the minimal cluster with CTEP despite the XES \CSs{} are considered on the central atom of all clusters used. One should also take into account that when increasing the cluster size, opportunities of its accurate treatment are dramatically diminishing because of problems  both with the basis set completeness and correlation treatment quality on the wave-function level.

For an additional comparison of the cluster calculations with periodic ones, the Nb-centered cluster was calculated with the same parameters as in calculation with the {\sc crystal} package (lesser basis set and 
all the pseudopotentials 
are used in spin-averaged approximation; the spin-orbit effects are taken into account in both cases only at final, one-center restoration stage \cite{Titov:05a} when calculating XES \CSs{}).
The XES \CSs{} on Nb in calculations of the clusters relative to crystal 
are presented in Table \ref{table:chemshifts-crystal-comparison}. As one can see, the difference between the periodic and cluster results does not exceed the overall errors of \CSs{} estimate 
   and, thus, can not serve as a basis for inferences.

\begin{table}[h!]
	\caption{Difference of X-ray chemical shifts on the Nb atom between cluster calculations and the periodic one (in meV)}
	\label{table:chemshifts-crystal-comparison}
	\begin{tabular}{lcc}
		\hline
		Structure & K$_{\alpha2, \alpha1}$ & K$_{\beta2, \beta1}$  \\  
		\hline 
		CTEP & 3 & 9 \\ 
		\hline 
		stoichiometric$\times1$ & -1  & 0 \\ 
		\hline 
		stoichiometric$\times4$ & -7  & -2 \\ 
		\hline 
		stoichiometric$\times8$ & 1  & 9 \\ 
		\hline
	\end{tabular}
\end{table}

\section{Conclusions}

A new method to simulate a fragment of ionic-covalent crystals within the cluster model, {\it compound-tunable embedding potential}, CTEP, is proposed. The CTEP method is based on modeling the embedding  potential by linear combination of short-range ``electron-free'' spherical  pseudopotentials for cations composing nearest environment of the main cluster, whereas the long-range CTEP part consists of only Coulomb potentials from environmental atoms. The short-range CTEP pseudopotentials for cations of nearest environment are {\it tuned} for the given crystal (on the basis of self-consistent semilocal pseudopotential version \cite{Titov:99}).
The long-range CTEP consists of Coulomb potentials from point charges centered on environmental atoms which are optimized as real numbers, positive for cations and  negative for anions. The total (electronic and nuclear) charge of the extended cluster (main cluster and nearest environment) is fixed as zero. The electronic structure relaxation of both main cluster and
   boundary regions
is taken into account within CTEP when one consider processes and point defects localized on the main cluster.  

A pilot application of the CTEP method to the fersmite crystal, CaNb$_2$O$_6$,
is performed and a remarkable agreement of the electronic density and optimized interatomic distances on the main cluster with the original crystal ones are attained. Local properties, i.e.\ characteristics of ``atoms-in-compounds'' \cite{Titov:14a} of primary interest for compound of $f$- and $d$-elements (Nb in fersmite), are considered on examples of chemical shifts of $K_{{\alpha}_{1,2}}$ and $K_{{\beta}_{1,2}}$ lines ($2p_{3/2,1/2}{\to}1s_{1/2}$ and $3p_{3/2,1/2}{\to}1s_{1/2}$, correspondingly) of X-ray emission (fluorescence) spectra.

This approach seems us promising for studying properties of point defects in solids, vacancies and impurities containing $f$- and heavy $d$-elements with relativistic effects and distortion of the crystal symmetry taken into account.
Application of the approach to consideration 
of adsorption of  superheavy elements on surfaces, effects of ionizing X-ray radiation, localized vibrations and rotations, magnetic structure of impurities of \dfel elements in materials, studying new physics in ferroelectrics, etc., is in progress.

\section{Acknowledgement}

This study was supported by the Russian Science Foundation (Grant No.~14-31-00022). We are grateful to I.V.\ Abarenkov, R.V.\ Bogdanov, S.G.\ Semenov and A.V.\ Zaitsevskii for many fruitful discussions.
Calculations in the paper were carried out using resources of the collective usage centre ``Modeling and predicting properties of materials'' at NRC “Kurchatov Institute” - PNPI.

\bibliography{bib/JournAbbr,bib/QCPNPI,bib/TitovLib}
\end{document}